\newcommand{\klesssim}{\mathrel{\hbox{\rlap{\hbox{\lower4pt\hbox{$\sim$}}}\hbox{$<$}}}}
\newcommand{\mpp}{m_{\rm p}}
\newcommand{\mcc}{m_{\rm c}}
\newcommand{\pb}{P_{\rm b}}

\documentclass{iau}

\title[IAUS291.~~Probing gravitation with pulsars] 
{Probing gravitation with pulsars} 

\author[M.~Kramer]  
{Michael Kramer$^{1,2}$}

\affiliation{$^1$MPI f\"ur Radioastronomy, Auf dem H\"ugel 69, 53121
  Bonn, Germany \\ email: {\tt mkramer@mpifr.de} \\[\affilskip]
$^2$Jodrell Bank Centre for Astrophysics, University of Manchester\\
Oxford Road, Manchester M13 9PL, UK \\email: {\tt michael.kramer@manchester.ac.uk}}

\pubyear{2012}
\volume{291}  
\jname{\mbox{Neutron Stars and Pulsars: Challenges and Opportunities after 80 years}}
\editors{J. van Leeuwen, ed.} 
\begin{document}

\maketitle

\begin{abstract}
Radio pulsars are fascinating and extremely useful objects. Despite
our on-going difficulties in understanding the details of their
emission physics, they can be used as precise cosmic clocks in a
wide-range of experiments -- in particular for probing gravitational
physics. While the reader should consult the contributions to these
proceedings to learn more about this exciting field of discovering, exploiting and
understanding pulsars, we will concentrate here on on the usage of
pulsars as gravity labs. 
\keywords{gravitation, techniques:radio astronomy, stars:neutron,  pulsars:general}
\end{abstract}


\firstsection 
\section{Introduction}

We are less than three years away from celebrating the centenary of
Einstein's theory of general relativity (GR). Nearly a hundred years
later, efforts in testing GR and its concepts are still being made by
many colleagues around the world, using many different approaches.  To
date GR has passed all experimental and observational tests with
flying colours, but in light of recent progress in observational
cosmology in particular, the question of as to whether alternative
theories of gravity need to be considered is as topical as ever.

Many experiments are designed to achieve ever more stringent tests by
either increasing the precision of the tests or by testing different,
new aspects.  Some of the most stringent tests are obtained by
satellite experiments in the solar system, providing exciting limits
on the validity of GR and alternative theories of gravity. However,
solar-system experiments are made in the gravitational weak-field
regime, while deviations from GR may appear only in strong
gravitational fields. 

We are all very much looking forward to the first direct detection of
gravitational waves with ground-based (and hopefully, eventually,
space-based) detectors, which not only open a completely new window to the
Universe but which will also provide superb tests of GR. Meanwhile, it happens
that nature provides us with an almost perfect laboratory to test the
strong-field regime - in the form of binary radio pulsars.

While, strictly speaking, the binary pulsars move in the weak
gravitational field of a companion, they do provide precision tests of
the strong-field regime. This becomes clear when considering that the
majority of alternative theories predicts strong self-field effects
which would clearly affect the pulsars' orbital motion. By
"simply" measuring the arrival time of pulsars moving in the curved
space-time of their companion, we can locate their position in their
orbit with an uncertainty as little as 20-50 m! Hence, tracing their
fall in a gravitational potential, we can search for tiny deviations
from GR, providing us with unique precision strong-field tests of
gravity.

\section{Simple and clean experiments}

The use of pulsars for tests of gravitational physics with ``pulsar
timing'' is a clean and conceptually simple experiment. By simply
measuring the exact arrival time of pulses at our telescope on Earth,
we do a ranging experiment that is vastly superior in precision than a
simple measurement of Doppler-shifts in the pulse period. In fact, the
pulsed nature of our signal that links tightly and directly to the
rotation of the neutron star, allows us to count every single rotation
of a neutron star (see contribution by Scott Ransom for details). In
this experiment we can consider the pulsar as a test mass that has a
precision clock attached to it and which allows us to follow the
movement of the test mass in the curved space-time of the companion.
As a result, a wide
range of relativistic effects can be observed, identified and studied.
These include so far:
\begin{itemize}
\item Precession of periastron
\item Gravitational redshift
\item Shapiro delay due to curved space-time
\item Gravitational wave emission
\item Geodetic precession, relativistic spin-orbit coupling
\item Speed of gravity
\end{itemize}
But we can also convert our observations in the tests of concepts and
principles deeply embedded in theoretical frameworks, such as
\begin{itemize}
\item Strong Equivalence Principle (grav. Stark effect),
\item Lorentz invariance of gravitational interaction,
\item Non-existence of preferred frames,
\item Conservation of total momentum,
\item Non-variation of gravitational constant,
\end{itemize}
which also leads to stringent limits on alternatives theories of
gravity, e.g. tensor--scalar theories, Tensor-Vector-Scalar (TeVeS) 
theories.

\subsection{Pick your laboratory!}

The various effects or concepts to be tested require sometimes a 
rather different type of laboratory. For instance, to test the
violation of the Strong-Equivalence-Principle, one would
like to use a binary system that consists of different types of masses
(i.e. with different gravitational self-energy), rather than a system
made of very similar bodies. Fortunately, nature has been kind.

At the moment, we know about 2000 pulsars, with about 10\% of these in 
binary systems. The shortest orbital period is about 90 min 
while the longest period is 5.3 yr (e.g.~Lorimer \& Kramer 2005). We find
different types of components, i.e.~main-sequence stars, white dwarfs
(WD), neutron stars (NS) and even planets. Unfortunately, despite 
past and on-going efforts, we have not yet found a pulsar about
a stellar black hole companion or about the supermassive black hole in the
centre of our Galaxy (Liu et al. 2012)\nocite{lwk+12}. Double neutron star systems are
rare but usually produce the largest observable relativistic effects
in their orbital motion and, as we will see, produce the best tests of
general relativity for strongly self-gravitating bodies. In
comparison, pulsar - white dwarf systems are much more common.
 Indeed, most pulsar companions are white dwarfs, with a wide range of
orbital periods, ranging from hours to days, weeks and months. Still,
many of them can be used for tests of gravitational theories where we
utilize the fact that white dwarfs and neutron stars differ very
significantly in their structure and, consequently, self-energies.

\subsection{Not only clean, but also very precise}

We follow the pulsars in their orbit by registering the variation in
arrival time as they move around the system's centre of mass. Using
pulsar timing techniques, we make extremely precise measurements
that allow us to probe gravitation with exquisite accuracy. Table 1
gives an idea about the precision that we already achieve today. With
future telescopes like the ``Square Kilometre Array'', the precision
will even be enhanced by at least two orders of magnitudes and should,
for instance, allow us to find a pular orbiting SGR A*, which would
provide the mass of the central black hole to a precision of an amazing
$1M_\odot$!  It would also allow us to measure the spin
of the black hole with a precision of $10^{-4}$ to $10^{-3}$ (enabling
tests of the ``cosmic censorship conjecture'') and the quadrupole moment
with a precision to  $10^{-4}$ to $10^{-3}$ (thus enabling tests of
the ``no-hair theorem''). See Liu et al. (2012) \nocite{lwk+12} and the 
contribution by Wex et al. for more details.

\begin{table}
\caption{Examples of precision measurements using pulsar timing as a
  demonstration what is possible today. The digit in bracket indicates
  the uncertainty in the last digit of each value. References are
  cited. More details on gravitational wave detection with pulsar
  timing arrays can be found in the contributions by Hobbs, Liu,
  Shannon and others.}

\begin{tabular}{lll}
\hline
Masses: & & \\
Masses of neutron stars: &    $m_1 = 1.4398(2) \,M_\odot$  &  (Weisberg et al. 2010) \\ 
 &   $m_2 = 1.3886(2) \,M_\odot$ &  (Weisberg et al. 2010) \\ 
Mass of WD companion:     &          		$0.207(2) \,M_\odot$ & (Hotan et al. 2006) \\
Mass of millisecond pulsar:	&	$1.67(2) \,M_\odot$ &(Freire et al. 2010) \\
Main sequence star companion:	&	$1.029(8) \,M_\odot$ & (Freire
et al. 2010) \\
Mass of Jupiter and moons:               &
$9.547921(2) \times 10^{-4} \,M_\odot$ &      	(Champion et al. 2010)\\
\noalign{\medskip}  
Spin parameters: & & \\
Period:	&	5.757451924362137(2) ms  & (Verbiest et al. 2008)  \\
\noalign{\medskip}
Orbital parameters: & & \\
Period:		&		$0.102251562479(8)$ day &    (Kramer
et al. in prep.) \\
Eccentricity:	&		$3.5 (1.1) \times 10^{-7}$ &
(Freire et al. in 2012) \\
\noalign{\medskip} 
Astrometry:& & \\
Distance:      &         	 		$157(1)$ pc	&	(Verbiest et al. 2008)\\
Proper motion: &  		    	$140.915(1)$ mas yr$^{-1}$ & 	(Verbiest e t al. 2008)\\
\noalign{\medskip} 
Tests of general relativity: & & \\
Periastron advance:	&		$4.226598(5)$ deg yr$^{-1}$ & 	(Weisberg et al. 2010)\\
Shrinkage due to GW emission: 	&                      $7.152(8)$
mm/day & 	(Kramer et al. in prep) \\
GR validity (obs/exp):	&	$1.0000(5)$ & 	(Kramer et al. in
prep.) \\
Constancy of grav. Constant, $\dot{G}/G$: &           $9(12) \times
10^{-13}$ yr$^{-1}$ &               (Zhu et al. in prep)\\
\noalign{\medskip}  
Gravitational wave detection: & & \\
Change in relative distance:      &                     100m / 1 lightyear       &    (EPTA, NANOGrav,  PPTA)\\
\hline
\end{tabular}
\end{table}

\section{The first laboratory}

The first binary pulsar to ever be discovered happend to be a rare
double neutron star system. It was discovered by Russel Hulse and Joe
Taylor in 1974 (Hulse \& Taylor 1975)\nocite{ht75a}. The pulsar, B1913+16, has a period of
59\,ms and is in eccentric ($e=0.61$) orbit around a unseen companion
with an orbital period of less than 8 hours. It became soon clear that
the pulsar does not follow the movement expected from a simple
Keplerian description of the binary orbit, but that it shows the
impact of relativistic effects.  In order to describe the relativistic
effects in a theory-independent fashion, one introduces so-called
``Post-Keplerian'' (PK) parameters that are included in a timing model
to accurately describe the measured pulse times-of-arrival (see
e.g.~contribution by David Nice for more details).

For the Hulse-Taylor pulsar, it was soon measured
that the system showed a relativistic advance of its periastron,
comparable to what is seen in the solar system for Mercury, albeit
with a much larger amplitude of $\dot\omega = 4.226598 \pm 0.000005$
deg/yr (Weisberg et al.~2010). GR predicts a value for the periastron advance that depends on
the Keplerian parameters and the masses of the pulsar and its
companion:
\begin{equation}
\dot{\omega} = 3 T_\odot^{2/3} \; \left( \frac{\pb}{2\pi} \right)^{-5/3} \;
               \frac{1}{1-e^2} \; (\mpp +
               \mcc)^{2/3}. \label{omegadot}
\end{equation}
Here, $T_\odot$ is a constant, $\pb$ the orbital period, $e$ the
eccentricity, and $\mpp$ and $\mcc$ the masses of the pulsar and its
companion. See Lorimer \& Kramer (2005)\nocite{lk05} for further details.

The Hulse-Taylor pulsars also shows the effects of gravitational redshift (including
a contribution from a second-order Doppler effect) as the pulsar moves
in its elliptical orbit at varying distances from the companion and
with varying speeds.  The result is a variation in the clock rate of
with an amplitude of $\gamma = 4.2992 \pm 0.0008$ ms (Weisberg et al.~2010). In GR, the
observed value is related to the Keplerian parameters and the masses as
\begin{equation}
\gamma  = T_\odot^{2/3}  \; \left( \frac{\pb}{2\pi} \right)^{1/3} \;
              e\frac{\mcc(\mpp+2\mcc)}{(\mpp+\mcc)^{4/3}}.
\end{equation}
We can now combine these measurements. We have two equations with a
measured left-hand side. On the right-hand side, we measured
everything apart from two unknown masses. We solve for those and
obtain, $\mpp = 1.4398 \pm 0.0002 \,M_\odot$ and $\mcc = 1.3886 \pm 0.0002
\,M_\odot$ (Weisberg et al. 2010).  These masses are correct if GR is the right theory of
gravity. If that is indeed the case, we can make use of the fact that
(for point masses with negligible spin contributions), the PK
parameters in each theory should only be functions of the a priori
unknown masses of pulsar and companion, $\mpp$ and $\mcc$, and the
easily measurable Keplerian parameters (Damour \& Taylor 1992)\nocite{dt92}\footnote{For
  alternative theories of gravity this statement may only be true for
  a given equation-of- state.}.  With the two masses now being
determined using GR, we can compare any observed value of a third PK
parameter with the predicted value. A third such parameter is the
observed decay of the orbit which can be explained fully by the emission
of gravitational waves. And indeed, using the derived masses, and the
prediction of general relativity, i.e.
\begin{equation}
\dot{P}_{\rm b} = -\frac{192\pi}{5} T_\odot^{5/3} \; \left( \frac{\pb}{2\pi} \right)^{-5/3} \;
               \frac{\left(1 +\frac{73}{24}e^2 + \frac{37}{96}e^4 \right)}{(1-e^2)^{7/2}} \; 
               \frac{\mpp \mcc}{(\mpp + \mcc)^{1/3}},
\end{equation}
one finds an agreement with the observed value of 
$\dot{P}_{\rm b}^{\rm obs} = (‐2.423 \pm 0.001) \times 10^{-12}$ 
(Weisberg et al.~2010) --
however, only if a correction for a relative acceleration between the
pulsar and the solar system barycentre is taken into account. As
the pulsar is located about 7 kpc away from Earth, it experiences a
different acceleration in the Galactic gravitational potential than
the solar system (see e.g. Lorimer \& Kramer 2005\nocite{lk05}). The precision of our knowledge to correct for this
effect eventually limits our ability to compare the GR prediction to
the observed value. Nevertheless, the agreement of observations and
prediction, today within a 0.2\% (systematic) uncertainty (Weisberg
et al.~2010)\nocite{wnt10}, represented the first evidence for the existence of
gravitational waves. Today we know many more binary pulsars where
we can detect gravitational wave emission. In one particular case, the
measurement uncertainties are not only more precise, but also the
systematic uncertainties are much smaller, as the system is much more
nearby. This system is the Double Pulsar.

\section{The Double Pulsar}

The Double Pulsar was discovered in 2003 (Burgay et al.~2003, Lyne et al.~2004)\nocite{bdp+03,lbk+04}. It
does not only show larger relativistic effects and is much closer to
Earth (about 1 kpc) than the Hulse-Taylor pulsar, allowing us to
largely neglect the relative acceleration effects, but the defining unique
property of the system is that it does not consist of one active
pulsar and its {\em unseen} companion, but that it harbours two {\em
  active} radio pulsars.

One pulsar is mildly recycled with a period of 22 ms (named ``A''), while the other
pulsar is young with a period of 2.8 s (named ``B''). Both orbit the common centre
of mass in only 147-min with orbital velocities of 1 Million km per
hour. Being also mildly eccentric ($e=0.09$), the system is an ideal
laboratory to study gravitational physics and fundamental physics in
general. A detailed account of the exploitation for gravitational
physics has been given, for instance, by Kramer et al.~(2006),
Kramer \& Stairs (2008) and Kramer \& Wex (2009)\nocite{ksm+06,ks08,kw09}. 
An update on those results is in preparation (Kramer et
al., in prep.), with the largest improvement undoubtedly given by a
large increase in precision when measuring the orbital decay. Not
even ten years after the discovery of the system, the Double Pulsar
provides the best test for the accuracy of the gravitational
quadrupole emission prediction by GR far below the 0.1\% level.

In order to perform this test, we first determine the mass ratio of
pulsar A and B from their relative sizes of the orbit, i.e.~$R=x_B/x_A
= m_A / m_B = 1.0714\pm 0.0011$ (Kramer et al.~2006). Note that this
value is theory-independent to the 1PN level (Damour \& Deruelle 1986). \nocite{dd86}
The most precise PK parameter that can be measured is a large orbital
precession, i.e.~$\dot\omega = 16.8991 \pm 0.0001$ deg/yr. 
Using Eq.~(\ref{omegadot}), this measured value and the mass ratio,
we can determine the masses of the pulsars, assuming GR is correct,
to be $m_A = (1.3381\pm0.0007) \,M_\odot$ and $m_B= (1.2489\pm0.0007)
\,M_\odot$.

We can use these masses to compute the expected amplitude for the
gravitational redshift, $\gamma$, if GR is correct. Comparing the result
with the observed value of $\gamma = 383.9 \pm 0.6\mu$s, we find that 
theory (GR) agrees with the observed value to a ratio of
$1.000\pm0.002$, as a first of five tests of GR in the Double Pulsar.

The Double Pulsar also has the interesting feature that the orbit is 
seen nearly exactly edge-on. This leads to a 30-s long eclipse of
pulsar A due to the blocking magnetosphere of B that we discuss further below,
but it also leads to a ``Shapiro delay'': whenever the pulse needs to
propagate through curved space-time, it takes a little longer than
travelling through flat space-time. At superior conjunction, when
the signal of pulsar A passes the surface of B in only 20,000km
distance, the extra path length due to the curvature of space-time
around B leads to an extra time delay of about 100$\mu$s. The
shape and amplitude of the corresponding Shapiro delay curve yield two
PK parameters, $s$ and $r$, known as {\em shape} and {\em range},
allowing two further tests of GR.
$s$ is measured to $s = \sin(i)=0.99975 \pm 0.00009$ and is in
agreement with the GR prediction of
\begin{equation}
s = T_\odot^{-1/3} \; \left( \frac{\pb}{2\pi} \right)^{-2/3} \; x \;
              \frac{(m_A + m_B)^{2/3}}{m_B} \label{eqn:s},
\end{equation}
(where $x$ is the projected size of the semi-major axis measured in lt-s)
 within a ratio of $1.0000\pm 0.0005$.
It corresponds to an orbital inclination angle of $88.7\pm0.2$ deg,
which is indeed very close to 90 deg as suggested by the eclipses.
$r$ can be measured with much less precision and yields an agreement
with GR's
value given by
\begin{equation}
r = T_\odot m_B, \label{eqn:r}\\
\end{equation}
to within a factor of $0.98\pm0.02$.

A fourth test is given by comparing an observed orbital decay of
$107.79\pm0.11$ ns/day to the GR prediction. Unlike the Hulse-Taylor
pulsar, extrinsic effects are negligible and the values agree with
each other without correction to within a ratio of
$1.000\pm0.001$. This is already a better test for the existence of GW
than possible with the Hulse-Taylor pulsar and will continue to improve
with time. Indeed, at the time of writing the
agreement has already surpassed the 0.1\% level significantly (Kramer et
al., in prep.).

\section{Relativistic spin-orbit coupling}

Apart from the Shapiro-delay, the impact of curved space-time is also
immediately measurable by its effect on the orientation of the pulsar
spin in a gyroscope experiment. This effect, known as geodetic
precession or de Sitter precession represents the effect on a vector
carried along with an orbiting body such that the vector points in a
different direction from its starting point (relative to a distant
observer) after a full orbit around the central
object. Experimental verification has been achieved
by precision tests in the solar system, e.g. by Lunar Laser Ranging
(LLR) measurements, or recently by measurements with the Gravity
Probe-B satellite mission (see Will~2006\nocite{wil06} for a review of
experimental tests). However, these tests are done in the weak field
conditions of the solar system, so that pulsars provide the only
access to the strong-field regime.

In binary systems one can interpret the observations, depending on the
reference frame, as a mixture of different contributions to
relativistic spin-orbit interaction. One contribution comes from the
motion of the first body around the centre of mass of the system
(deSitter-Fokker precession), while the other comes from the dragging
of the internal frame at the first body due to the translational
motion of the companion (B\"orner et al.~1975)\nocite{ber75}. 
Hence, even though we loosely talk about
geodetic precession, the result of the spin-orbit coupling for binary
pulsar is more general, and hence we will call it {\em relativistic
  spin-precession}. The consequence of relativistic spin-precession is a precession of
the pulsar spin about the total angular moment vector, changing the
orientation of  the pulsar relative to Earth.

Since the orbital angular momentum is much larger than the angular
momentum of the pulsar, the orbital spin practically represents a
fixed direction in space, defined by the orbital plane of the binary
system. Therefore, if the spin vector of the pulsar is misaligned with the
orbital spin, relativistic spin-precession leads to a change in
viewing geometry, as the pulsar spin precesses about the total angular
momentum vector.  Consequently, as many of the observed pulsar
properties are determined by the relative orientation of the pulsar
axes towards the distant observer on Earth, we should expect a
modulation in the measured pulse profile properties, namely its shape
and polarisation characteristics (Damour \& Ruffini 1974).\nocite{dr74}
The precession rate is another PK parameter and given in GR by
(e.g.~Lorimer \& Kramer 2005)
\begin{equation}
\label{eqn:om}
\Omega_{\rm p}  =  T_\odot^{2/3} \times \left( \frac{2\pi}{P_{\rm b}}\right)^{5/3} \times
 \frac{m_{\rm c}(4m_{\rm p}+3m_{\rm c})}{2(m_{\rm p}+m_{\rm c})^{4/3}} \times
 \frac{1}{1-e^2}
\end{equation}
In order to see a measurable effect in any binary pulsar, {\em a)} the spin
axis of the pulsar needs to be misaligned with the total angular
momentum vector and {\em b)} the precession rate must be sufficiently
large compared to the available observing time to detect a change in
the emission properties. Table 2 lists the known Double
Neutron Star Systems which typically show the largest degree of
relativistic effects due to the often short eccentric binary
orbits. However, the last entry in the table is PSR J1141$-$6545 which
is a relativistic system with a white dwarf companion. 
Those pulsars that are marked with an asterisk have been identified as
pulsars showing relativistic spin precession. Note that the top 5 out of 8 sources
(with a known expected precession rate) indeed show the
effect.

As the most relativistic binary system known to date, we expect a
large amount of spin precession in the Double Pulsar system.  Despite
careful studies, profile changes for A have not been detected,
suggesting that A's misalignment angle is rather small (e.g.~Ferdman
et al.~2012)\nocite{fsk+12}. In contrast, changes in the light curve and pulse
shape on secular timescales (Burgay et al.~2005)\nocite{bpm+05} reveal that this is not the
case for B. In fact, B had been becoming progressively weaker and
disappeared from our view in 2009 (Perera et al.~2010)\nocite{pmk+10}. Making the valid
assumption that this disappearance is solely caused by relativistic
spin precession, it will only be out of sight temporarily until it
reappears later. Modelling suggests that, depending on the beam shape,
this will occur in about 2035 but an earlier time cannot be
excluded. The geometry that is derived from this modelling is
consistent with the results from complementary observations of spin
precession, visible via a rather unexpected effect described in the
following.

The change on the orientation of B also changes the observed eclipse
pattern in the Double Pulsar, where we can see periodic bursts of
emission of A during the dark eclipse phases, with the period being
the full- or half-period of B. As this pattern is caused by the
rotation of B's blocking magnetospheric torus that allows light to
pass B when the torus rotates to be seen from the side, the resulting
pattern is determined by the three-dimensional orientation of the
torus, which is centred on the precessing pulsar spin. Eclipse
monitoring over the course of several years shows exactly the expected
changes, allowing to determine the precession rate to $\Omega_{\rm p,
  B}= 4.77^{+0.66}_{-0.65}$ deg/yr.  This value is fully consistent
with the value expected GR, providing a fifth test (Breton et
al.~2008).\nocite{bkk+08} 
This measurement also allows to test alternative theories of gravity and
their prediction for relativistic spin-precession in strongly
self-gravitating bodies for the first time (see Kramer \& Wex 2009 for
details).

\section{Alternative theories}

Despite the successes of GR, a range of observational data have fuelled
the continuous development of alternative theories of gravity.  Such
data include the apparent observation of ``dark matter'' or the
cosmological results interpreted in the form of ``inflation'' and
``dark energy''. Confronting alternative theories with data also in
other areas of the parameter space (away from the CMB or Galactic
scales), requires that these theories are developed sufficiently in
order to make predictions. A particular sensitive criterion is if the
theory is able to make a statement about the existence and type of
gravitational waves. Most theories cannot (yet), but a class of
theories where this has been achieved is the class of tensor-scalar
theories as discussed and demonstrated by Damour and Esposito-Far{\`e}se
in a series of works (e.g. Damour \& Esposito-Far{\`e}se 1996)\nocite{de96}.  For corresponding tests, the
choice of a double neutron star system is not ideal, as the difference
in scalar change, (that would be relevant, for instance, for the
emission of gravitational {\em dipole} radiation) is small. The ideal
laboratory would be a pulsar orbiting a black hole, as the black hole
would have zero scalar charge and the difference would  be maximised. The
next best laboratory is a pulsar-white dwarf system. Indeed, such binary
systems are able to provide constraints for alternative theories of
gravity that are equally good or even better than solar system limits
(Freire et al.~2012). \nocite{fwe+12}

\begin{table}
\label{tab2}
\caption{DNSs sorted according to the expected relativistic spin precession rate.
  Also included is PSR   J1141$-$6545 which is in a 
  relativistic orbit about a white dwarf companion. Pulsars marked
  with an asterisk have been identified of showing spin precession.
 For sources where no precession rate is listed,
  the companion mass could not be accurately measured yet, indicating 
  however, that the precession rate is low.}

\begin{center}
{\begin{tabular}{lrcccc}
    \hline
    PSR & $P$(ms) & $P_{\rm b}$ (d) &  $x$(lt-s) & $e$& $\Omega_{\rm p}$ (deg yr$^{-1}$) \\
    \hline
    J0737$-$3039A/B$^\ast$ & 
    \multicolumn{1}{c}{22.7/2770} & 0.10 & 1.42/1.51 & 0.09 & \multicolumn{1}{c}{4.8/5.1} \\
    J1906+0746$^\ast$ & 144.1 & 0.17 & 1.42 & 0.09 & 2.2 \\
    B2127+11C$^\ast$ & 30.5 & 0.34 & 2.52 & 0.68 & 1.9 \\
   B1913+16$^\ast$ & 59.0 & 0.33 & 2.34 & 0.62 & 1.2 \\
    J1756$-$2251 & 28.5 & 0.32 & 2.76 & 0.18 & 0.8 \\
    B1534+12$^\ast$ & 37.9 & 0.42 & 3.73 & 0.27	& 0.5 \\
   J1829+2456 & 41.0 & 1.18 & 7.24 & 0.14 & 0.08 \\
    J1518+4904 & 40.9 & 8.64 & 20.0 & 0.25 & -- \\
   J1753$-$2240 & 95.1 & 13.63 & 18.1 & 0.30 & -- \\
   J1811$-$1736 & 104.2 & 18.8 & 34.8 & 0.83 & -- \\
   J1141$-$6545$^\ast$ & 394.0 & 0.20 & 1.89 & 0.17 & 1.4 \\
    \hline
\end{tabular}}
\end{center}
\end{table}

\section{Summary \& Conclusions}

A variety of experiments and observational data exist that allow us to
test our understanding of gravity with increased precision. So far,
general relativity has passed all tests with flying colours but the
apparent existence of ``Dark Energy'' challenges this simple picture.
It is clear that the observations of pulsars will continue to play an
important part in testing general relativity and its alternatives. We
expect to detect gravitational waves not only indirectly but also
directly using pulsar observations (see corresponding contributions),
and we have all reasons to believe that future searches will yield
pulsars that can probe the space-time around black holes. Combined with
the results of other experiments, namely the detection of
gravitational waves with ground based detectors, we can expect a
bright future for our understanding of gravity.

%
%

\end{document}